# Proximity-induced interfacial room-temperature ferromagnetism in semiconducting $Fe_3GeTe_2$

Qianwen Zhao[1,2], Yingmei Zhu[3,4], Hanying Zhang[1,2], Baiqing Jiang[1,2], Yuan Wang[1,2], Tunan Xie[1,2], Kaihua Lou[1,2], ChaoChao Xia[1,5], Hongxin Yang[3,4*], and C. Bi[1,2,5*]

[1]Laboratory of Microelectronic Devices & Integrated Technology, Institute of Microelectronics, Chinese Academy of Sciences, Beijing 100029, China

[2]University of Chinese Academy of Sciences, Beijing 100049, China

[3]National Laboratory of Solid State Microstructures, School of Physics, Collaborative Innovation Center of Advanced Microstructures, Nanjing University, Nanjing 210093, China

[4]Ningbo Institute of Materials Technology and Engineering, Chinese Academy of Sciences, Ningbo 315201, China

[5]School of Microelectronics, University of Science and Technology of China, Hefei 230026, China

*hongxin.yang@nju.edu.cn
*clab@ime.ac.cn

# Abstract

The discoveries of two-dimensional ferromagnetism and magnetic semiconductors highly enrich the magnetic material family for constructing spin-based electronic devices but with an acknowledged challenge that the Curie temperature ($T_c$) is usually far below room temperature. Many efforts such as voltage control and magnetic ion doping are currently underway to enhance the functional temperature, in which the involvement of additional electrodes or extra magnetic ions limits their plenty of applications in practical devices. Here we demonstrate that the magnetic proximity, a robust effect but with elusive mechanisms, can induce room-temperature ferromagnetism at the interface between sputtered Pt and semiconducting $Fe_3GeTe_2$, both of which do not show ferromagnetism at 300 K. The independent electrical and magnetization measurements, structure analysis, and control samples with Ta highlighting the role of Pt confirm that the ferromagnetism with the $T_c$ of above 400 K arises from the $Fe_3GeTe_2$/Pt interfaces, rather than Fe aggregation or other artificial effects. Moreover, contrary to conventional ferromagnet/Pt structures, the spin current generated by the Pt layer is enhanced more than two times at the $Fe_3GeTe_2$/Pt interfaces, indicating the potential applications of the unique proximity effect in building high-efficient spintronic devices. These results may pave a new avenue to create room-temperature functional spin devices based on low-$T_c$ materials and provide clear evidences of magnetic proximity effects by using non-ferromagnetic materials.

**Keywords**

*magnetic proximity effects, spin-orbit torque, 2D ferromagnetism, Fe₃GeTe₂, interfacial magnetism, spin Hall magnetoresistance*

# 1. Introduction

Along with the rapid development of spintronic devices based on traditional magnetic materials like Fe, Co, Ni, and their alloys that have promoted the performance improvement of magnetic sensors and memories[1], novel spintronic materials have also been explored to construct more functional devices for broader applications[2–5]. One typical example is magnetic semiconductors, which were ignited by the idea taking the advantage of mature semiconductor technologies for the production of spintronic devices[2,6]. Unfortunately, the intrinsic ferromagnetism of pure semiconductors only exists below 110 K[7,8], making them not practical for applications above room temperature. Although extra measures such as magnetic ion doping or electrical control were claimed to increase their $T_c$[2,6,9], there are many concerns about the magnetic precipitates or contaminations that may contribute the reported weak ferromagnetism[10]. So far, the researches on the magnetic semiconductors are still in the stage of fundamental material characterizations and no products have been released. The same situation happens for another typical example, the recently discovered two-dimensional (2D) ferromagnets[11,12]. In 2D materials, the long-range ferromagnetic order was not expected in the past several decades because of thermal fluctuations[13] until it was demonstrated in the $Cr_2Ge_2Te_6$ and $CrI_3$ atomic layers recently, even though the corresponding $T_c$ is below 45 K in the monolayer limit[11,12]. Currently, many efforts have been devoted to the improvement of $T_c$, from exploring new layered material systems based on theoretical predictions to introducing extrinsic contributions like the measures adopted for magnetic semiconductors[14–21]. These extra control approaches such as electric field-driven ionic control[19–22] may provide new opportunities for fundamental spintronics, but the reliable ferromagnetism above room temperature in a simple structure which is facile to build spintronic devices has not been achieved. One particular example is that the interfacial exchange coupling induced by a topological insulator has been demonstrated to increase the $T_c$ of neighbored 2D layers[23,24].

The magnetic proximity effect (MPE) in which the magnetism can be induced in a nonmagnetic layer (NM) adjacent to a ferromagnet (FM) has been an active but debated research topic in recent years[25–31]. It may contribute to the interfacial spin Hall magnetoresistance (SMR)[32,33], thermoelectric Seebeck effects, and spin-orbit torques in NM/FM bilayers and even make 2D NM showing ferromagnetic behaviors[26,27,31–33]. Although most MPEs were investigated through electrical transport measurements by using a NM/YIG bilayer where YIG is a ferromagnetic insulator for excluding electrical transport from the FM itself, there are still many debates on the interfacial spin transport[31,33]. So far, only one fact is clear that the MPE-induced magnetic properties such as coercivity and saturation magnetic fields at the NM/FM interfaces strongly rely on the ferromagnetic properties of the FM[25–31], making the MPE originating from the FM itself more convincible. Therefore, it will be promising if the MPE can be demonstrated in a system in the absence of ferromagnetism, which may eliminate the MPE debates and also enrich the applications of NM, especially semiconductors and 2D materials, in spintronic devices. As schematically shown in Figure 1, here we choose such non-ferromagnetic systems and report the observation of MPE and resultant room-temperature ferromagnetism at the interface between two non-ferromagnetic materials, which can highly enhance the functional temperature of adjacent low-$T_c$ materials. Specifically, we have observed the ferromagnetic characteristics of the sputtered $Fe_3GeTe_2$/Pt (FGT/Pt) bilayers above the room temperature, where both the sputtered FGT and Pt single layers are not ferromagnetic at 300 K. In contrary to the previously reported MPEs, the observed room-temperature ferromagnetism does not depend on any FMs and completely arises from the proximity effects at the FGT/Pt interfaces. The remarkable MPE and room-temperature ferromagnetism were further examined by investigating spin current absorption at the FGT/Pt interfaces. On the other hand, from the point of view of practical applications, the demonstrated interfacial MPE provides a reliable approach to increase the functional temperature of those newly developed low-dimensional materials to room-temperature in the absence of magnetic ion doping or extra electrodes. Since both the amorphous and crystallize FGT show similar ferromagnetic properties[34], the MPE reported here may also persist in a crystallize FGT/Pt interface.

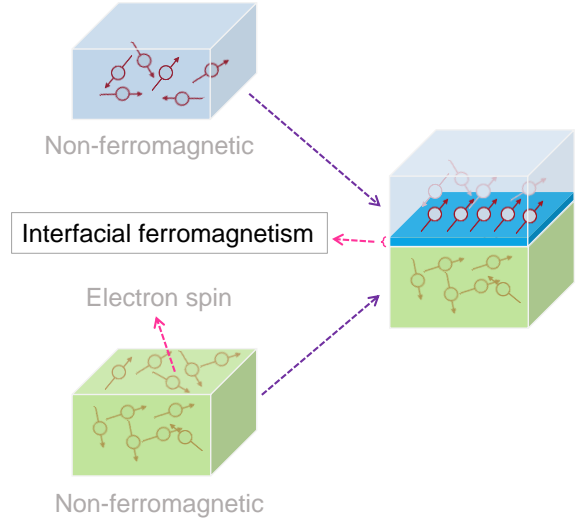

Figure 1. Schematic of proximity-induced interfacial ferromagnetism. Electron spin prefers parallel alignments due to possible enhanced exchange coupling at the interface, which results in the interfacial ferromagnetism.

## 2. Results and Discussion

The Sub/FGT ($t_{FGT}$)/Pt ($t_{Pt}$) and Sub/Pt ($t_{Pt}$)/FGT ($t_{FGT}$) multilayers with the capping layer of either 2.5 nm TaO$_x$ or 10 nm SiO$_2$ were deposited sequentially on the Si/SiO$_2$ (300 nm) substrate, where 3 nm ≤ $t_{FGT}$ ≤ 120 nm and 3 nm ≤ $t_{pt}$ ≤ 10 nm are the nominal thicknesses of the FGT and Pt layers, respectively. The control samples without Pt were also deposited for characterizing the structural and magnetic properties of the single FGT layers. As shown in Figure 2a, even for the 120 nm FGT single layer, there are no observable X-ray diffraction (XRD) peaks attributed to crystallized Fe$_3$GeTe$_2$, Fe, Ge, or Te, indicating that the sputtered FGT layers are amorphous and lack of 2D crystalline orders. For the FGT (3)/Pt (3) bilayers (unit: nm), a broad peak centered at 39.8° can be attributed to the Pt (111) peak. The crystallized Pt and amorphous FGT layers were also confirmed by the high-resolution transmission electron microscopy (HRTEM) analysis. As shown in Figure 2b, the crystal lattice in the Pt side can be observed clearly while no visible crystal lattice was found in the FGT region. The corresponding element mapping in Figure 2c further demonstrates that the Fe, Ge, and Te distribute in the entire FGT region and no aggregation of single element is formed. Particularly, there is no any crystallized Fe or Fe atom aggregation found in the FGT region as well as at the FGT/Pt interfaces, which are very important to exclude the possible origin of the following observed room-temperature ferromagnetism from Fe aggregation.

The basic electrical characteristics of sputtered FGT and FGT/Pt were investigated by using a Hall bar structure with the length of 200 μm and width of 10 μm. Contrary to the crystallized FGT showing metallic transport with the thickness larger than six monolayers[21], the sputtered FGT always show semiconducting behaviors, as evidenced in the Figure 2d, in which the resistance ($R$) reduces with increasing temperature for all FGT single layers with or without a capping layer (with an oxidized top surface if no capping layer). For the FGT (10)/Pt (3), the metallic transport indicates that the Pt layer dominates the electrical transport in FGT/Pt bilayers.

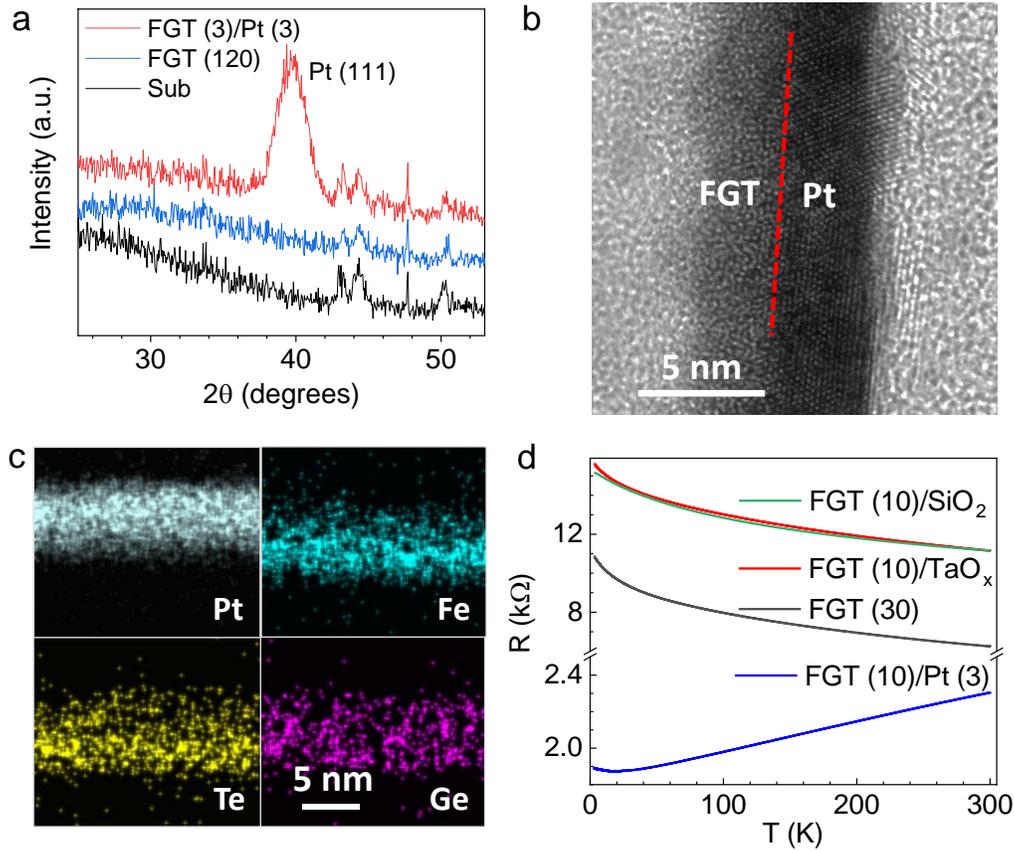

Figure 2. Structural and electrical characterization of FGT/Pt bilayers. a) XRD results of FGT (3)/Pt (3) and FGT (120). The small peaks between 42° and 52° arises from bare substrates. b) HRTEM images of FGT (3)/Pt (3) bilayers and c) corresponding element mapping. The red dash line indicates the FGT/Pt interface. d) The recorded resistance as a function of temperature for FGT (10)/Pt (3) and control samples with $SiO_2$, $TaO_x$, or no capping layers.

The Hall resistance ($R_H$) was directly measured to investigate the magnetic properties. The current (*I*) and external magnetic field (*H*) were applied along the *x* and *z* directions, respectively, as illustrated in the inset of Figure 3a. It is well known that $R_H$ shows a linear *H* dependence for non-ferromagnetic materials and an additional $R_s|\mathbf{M}|$ dependence for the ferromagnets with magnetization, **M**, and the anomalous Hall coefficient, $R_s$. The additional Hall contribution has been widely used to describe ferromagnetism both experimentally and theoretically[35]. As shown in Figure 3a, the room-temperature $R_H$ curves show a linear *H* dependence for the sputtered single FGT layers, which is the same as the crystallized FGT because of no ferromagnetism[18,21,34]. Remarkably, for the FGT (10)/Pt (3) bilayer, clear hysteresis behaviors appear in the $R_H$ curve at room temperature, indicating that the ferromagnetism has been induced in the FGT/Pt bilayer with $T_c$ larger than 300 K. Since both the FGT and Pt single layers do not show ferromagnetism at room temperature, the ferromagnetism must arise from the FGT/Pt interfaces due to proximity effects. For the Sub/Pt (3)/FGT (10)/capping layer where the Pt layer was deposited first, $R_H$ shows the same behaviors as FGT (10)/Pt (3) (Figure S3), indicating that the observed room-temperature ferromagnetism does not depend on the deposition sequence of the FGT and Pt layers. This excludes that the ferromagnetism may arise from the possible Fe aggregation due to the discrepancy of atomic mass or ion energy among sputtered Pt, Fe, Ge, and Te elements, the size of which may be less than the spatial resolution of HRTEM. In fact, the Fe aggregation with the size of several nanometers is superparamagnetic and should not show any magnetic anisotropy[36], which conflicts with the following anisotropic magnetoresistance (AMR) measurement results.

Figure 3b and 3c show the longitudinal resistance as a function of *H* measured at room temperature. For the FGT (10)/Pt (3) bilayers, the measured resistance strongly depends on the applied *H* as shown in Figure 3b, further confirming the presence of room-temperature ferromagnetism. In particular, two clear resistance peaks corresponding to the magnetization switching under an $H_x$, rather than $H_y$ and $H_z$, indicate an in-plane *shape* magnetic anisotropy in the FGT/Pt bilayer, which cannot be induced by superparamagnetic Fe aggregation. Moreover, in sharp contrary to the AMR mechanism with $R(H_z) \approx R(H_y) < R(H_x)$ when *H* is larger than the saturation field[37], Figure 3b shows $R(H_z) \approx R(H_x) > R(H_y)$ that obeys the spin Hall magnetoresistance (SMR) mechanism[32,33]. SMR arises because the spin current generated through the spin Hall effects (SHEs) in the Pt layer scatters at the surface by a neighbored ferromagnet[32]. In this case, the spin current is absorbed by the ferromagnet when the spin polarization (along the $y$ direction) and **M** of the ferromagnet is in a noncolinear configuration and reflected in a colinear configuration. The former results in a high resistance state (**M** ⊥ **y**) and the latter lead to a low resistance state (**M** ∥ **y**), in consistent with the FGT (10)/Pt (3) results[32,33]. The detailed angle dependent resistance measurement results will be presented later. The occurrence of SMR further reveal the presence of room-temperature ferromagnetism, rather than superparamagnetism, at the FGT/Pt interface. For the single FGT layer, as shown in Figure 3c, $R$ does not change with $H$ due to the lack of ferromagnetism as evidenced by the Hall measurements (Figure 3a). These magnetoresistance results agree well with corresponding Hall results for both FGT/Pt and single FGT. The ferromagnetism of FGT (10)/Pt (3) bilayers was also directly measured by using a vibrating sample magnetometer (VSM). As shown in Figure 3d, after subtraction of a linear background, the VSM curves measured under an in-plane field show typical ferromagnetic characteristics that are sustained even at 400 K.

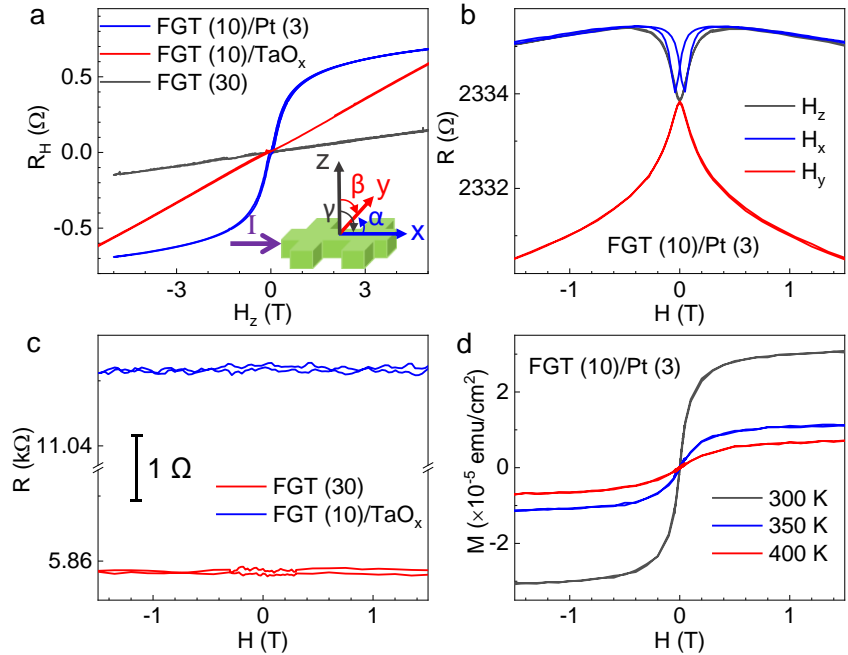

Figure 3. Ferromagnetism of FGT (10)/Pt (3) bilayers. a) $R_H$ as a function of $H_z$ for FGT (10)/Pt (3), FGT (10)/TaO$_x$, and FGT (30) at 300 K. The inset illustrates experimental configurations for electrical measurements. b, c) Magnetoresistance measurement results for (b) FGT (10)/Pt (3) bilayers and (c) FGT single layers at 300 K. d) Temperature-dependent magnetization hysteresis loops for FGT (10)/Pt (3) measured by VSM.

Figure 4 presents the temperature dependence of ferromagnetism measured by using both the Hall and VSM measurements. For the FGT (10)/Pt (3) bilayers, the ferromagnetic $R_H$ signal increases with decreasing temperature like most ferromagnets. To reveal the possible ferromagnetic contribution from the FGT itself, $R_H$ of the 30 nm FGT single layer was also measured. As mentioned above, the FGT single layer does not show ferromagnetic behaviors at room temperature, and the measured $R_H$ curves only show normal linear Hall signal above 200 K. As reported in previous works, even for the sputtered amorphous FGT lacking long-range crystalline order, when the temperature drops below 180 K, $R_H$ signal increases dramatically and becomes hysteretic with decreasing temperature[34]. The ferromagnetic transition temperature is the same as the well-crystallized counterparts with the $T_c$ of about 180 K[18,21]. To clearly demonstrate the temperature dependence of ferromagnetism, the anomalous Hall resistance ($R_H^S$) originating from ferromagnetism was extracted by subtracting a linear normal Hall contribution. Figure 4c shows the extracted $R_H^S$ as a function of temperature for both samples. One can see that $R_H^S$ of the single FGT drops much faster than that of the FGT/Pt with increasing temperature and finally to zero around 200 K, while for the FGT/Pt bilayer, a sizable $R_H^S$ persists at 300 K because of the ferromagnetism. Figure 4d gives the measured magnetization of the FGT (10)/Pt (3) bilayer by using VSM as a function of temperature under a 200 Oe in-plane field. Contrary to the temperature-dependent $R_H^S$ results, the magnetization drops quickly around 200 K, which is consistent with the transition temperature of $R_H^S$ for the single FGT (Figure 4c), indicating the main ferromagnetic contribution arising from the FGT layer below 200 K. Above 300 K, the magnetization decreases continuously with increasing temperature, where the magnetization is dominated by the ferromagnetism induced at the FGT/Pt interface. These results verify the ferromagnetic contribution from both the FGT/Pt interface and FGT itself varying with temperature and further confirm the proximity-induced interfacial ferromagnetism at room temperature.

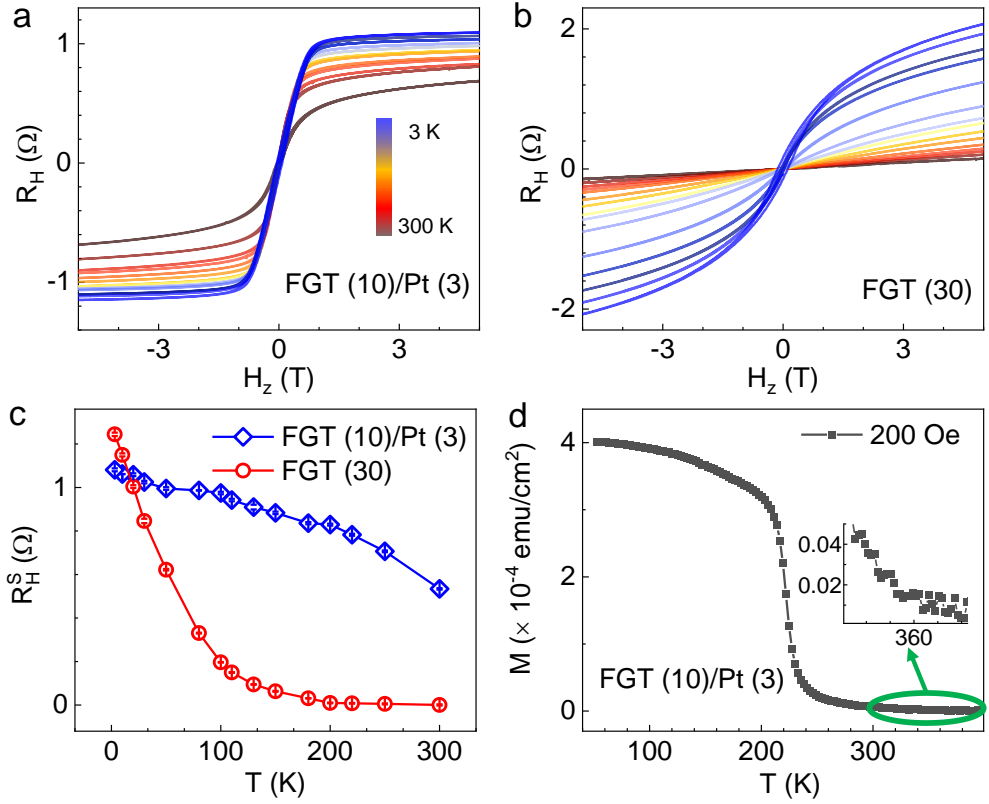

Figure 4. Temperature dependence of proximity-induced ferromagnetism. a, b) The recorded $R_H$ curves of (a) FGT (10)/Pt (3) and (b) FGT (30) from 3 K to 300 K. c) The extracted ferromagnetic contribution of $R_H$ from (a) and (b) as a function of temperature. d) Temperature-dependent magnetization measurement results with a 200 Oe in-plane magnetic field. The inset is the enlarged region around 350 K.

To further reveal the ferromagnetism and MPE at the FGT/Pt interface, the angle dependent resistance was measured under a 4 T magnetic field. As shown in Figure 3d and 4a, the 4 T magnetic field is large enough to align magnetization along the applied magnetic field direction. The AMR due to ferromagnetism itself can be evaluated by rotating $H$ in the $zx$ plane (γ scan) because the SHE-induced spin polarization (along the $y$ direction) keeps perpendicular to the magnetization in this plane and the resistance due to SMR does not change[32,33]. Similarly, the interfacial SMR can be evaluated by rotating $H$ in the $zy$ plane (β scan) where AMR keeps the same. Figure 5a-5c show the angle-dependent $R$ measured at 300 K, 100 K, and 3 K, respectively. Generally, all the angle-dependent curves obey a *sine* or *cosine* dependence, which can be explained within the AMR or SMR theories[32,33,37]. For the γ scan, the maximum *R* appears around 90° and 270° while the minimum *R* around 0° and 180°, which is the typical AMR behaviors for most ferromagnets with *R* (**M** ⊥ **I**) < *R* (**M** ∥ **I**)[37]. For the β scan, the minimum *R* is obtained when **M** is along the *y* direction, also consistent with the SMR mechanism. The resistance change in the α scan can be viewed as the combine of both AMR and SMR contributions. As shown in Figure 5a-5c, the AMR contribution to the resistance change in the α scan gradually increases with decreasing temperature, which can be explained by the enhancement of ferromagnetism as well as the ferromagnetic contribution from the FGT itself at low temperatures. The calculated AMR and SMR ratios, $\Delta R/R$, as a function of temperature are shown in Figure 5d, where *ΔR* is the maximum resistance change during angle scans. The monotonic decrease of AMR ratio with increasing temperature agrees with the typical temperature-dependent ferromagnetic behaviors, while the SMR ratio first increases and then decrease around 200 K, which is more complex compared to other Pt/ferromagnetic insulator (Pt/FI) systems[38,39]. In the Pt/FI bilayer, SMR increases with temperature and saturates around 100 K[38]. The SMR transition temperature around 200 K for the FGT/Pt bilayer corresponds to the $T_c$ of FGT layer, indicating that the ferromagnetism from both the FGT layer and FGT/Pt interface contribute to the SMR. Above 200 K, the FGT layer becomes paramagnetic and SMR arises from the interfacial ferromagnetism only. The decrease trend of SMR in this temperature range can be understood that the interfacial ferromagnetism gradually becomes weak with increasing temperature as demonstrated by VSM results in Figure 3d.

Moreover, the SMR ratio is about one time larger than the Pt/FI bilayers[38] (compared to some works[33,39], the SMR ratio reported in this work is even one order larger). This indicates that the spin current generated by the Pt layer is enhanced at the FGT/Pt interface. It should be noted that the calculated SMR ratio could be larger by excluding the current shunting effects due to the semiconducting FGT layer. According to SMR theories, SMR arises from the absorption of transverse spin current (spin polarization perpendicular to the magnetization of FM) by the neighbored FM[32,40]. The large SMR ratio corresponds a large spin Hall angle ($\theta_{SH}$) of the Pt layer even by considering a transparent FGT/Pt interface (the spin mixing conductance, $G_r \to \infty$, or $1 \times 10^{15}$ Ω$^{-1}$ m$^{-2}$ for calculation[40]). The SMR ratio can be written as

$$\frac{\Delta R}{R} \approx \theta_{SH}^2 \frac{\lambda_{Pt}}{t_{Pt}(1+\xi)} \frac{2\rho_{Pt}\lambda_{Pt}G_r \tanh^2\left(\frac{t_{Pt}}{2\lambda_{Pt}}\right)}{1+2\rho_{Pt}\lambda_{Pt}G_r \coth\left(\frac{t_{Pt}}{\lambda_{Pt}}\right)}, \quad (1)$$

where $\xi = \frac{\rho_{Pt} t_{FGT}}{\rho_{FGT} t_{Pt}}$ for correcting the current shunting of the FGT layer. $\lambda_{Pt}$, $\rho_{Pt}$ = 43.65 µΩ cm, and $\rho_{FGT}$ = 560.50 µΩ cm represent spin diffusion length, resistivity of the Pt layer, and resistivity of the FGT layer, respectively. By choosing a reasonable $\lambda_{Pt}$ = 1.03 nm, the low limit of estimated $\theta_{SH}$ is 0.097 at room temperature, which is about two times larger than those reported in Pt/FI systems by using the same characterization techniques[33,39]. It should be noted that, for a larger or smaller $\lambda_{Pt}$, or by considering the absorption of longitudinal spin current (spin polarization parallel to the magnetization of FM) like metallic bilayers[40], an even larger $\theta_{SH}$ is required to explain the measured SMR ratio. With $\theta_{SH}$ = 0.097, the calculated $\lambda_{Pt}$ as a function of temperature is shown in Figure S1 (Supporting Information), in which both the magnitude and temperature dependence (decreasing with increasing temperature) of $\lambda_{Pt}$ are reasonable and consistent with other reports[39]. The large $\theta_{SH}$ indicates that the FGT/Pt interface is indeed different from conventional Pt/FI interfaces. As demonstrated here, the proximity effect not only induces the room temperature ferromagnetism at the FGT/Pt interfaces but also results in an enhanced spin current. The spin memory loss[41] or interface roughness cannot account for the observed $\theta_{SH}$ since a transparent FGT/Pt interface ($G_r = 1 \times 10^{15}$ Ω$^{-1}$ m$^{-2}$) has been adopted for calculation. The spin current generated through the anomalous Hall effects or SHE of the interfacial ferromagnetism like a metallic FM[42,43] could be an interesting alternative source, which requires further investigation.

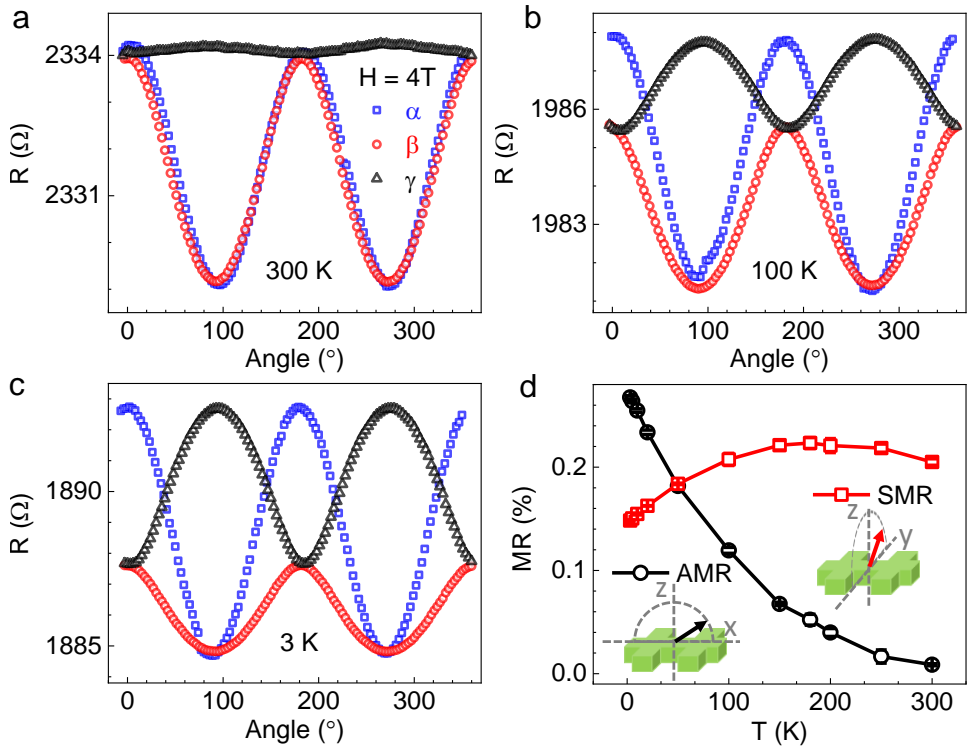

Figure 5. Temperature-dependent magnetoresistance for FGT (10)/Pt (3). a-c) Typical angle-dependent resistance measured at (a) 300 K, (b) 100 K, and (c) 3 K. The applied magnetic field was 4 T. d) The extracted AMR and SMR ratios as a function of temperature. The insets show schematic AMR and SMR measurements.

In Supplementary Information, we have calculated magnetic exchange coupling between two Fe ions at the crystalline FGT/Pt interfaces through the first-principles calculations. We found that the exchange coupling between $Fe^{3+}$ and $Fe^{2+}$ in the crystalline FGT monolayer can be enhanced about two times in the presence of Pt atoms near the Fe or Ge sites, which indicates that the exchange coupling may also be enhanced at amorphous FGT/Pt interfaces, leading to the observed room-temperature ferromagnetism.

## 3. Conclusions

In conclusion, we have observed proximity-induced room-temperature ferromagnetism sustaining even above 400 K at the FGT/Pt interfaces. The room-temperature ferromagnetism was confirmed by the Hall, magnetoresistance, and conventional VSM measurements independently. The AMR and SMR switching behaviors indicate that the room-temperature ferromagnetism has an in-plane anisotropy and arises from the FGT/Pt interface. Moreover, a large $\theta_{SH} = 0.097$, about two times larger than that in other Pt/FI systems, is obtained at the FGT/Pt interfaces. These results not only provide strong evidences for MPE, but also suggest a facile approach to construct room-temperature spintronic devices based on the non-ferromagnetic materials, especially for the ultrathin low-dimensional materials that are very sensitive to interfaces.

## 4. Experimental Section

*Sample fabrication*: FGT ($t_{FGT}$)/Pt ($t_{Pt}$) and Pt ($t_{Pt}$)/FGT ($t_{FGT}$) multilayers capped with TaO$_x$ (2.5 nm) or SiO$_2$ (10 nm) were deposited on thermally oxidized Si/SiO$_2$ (300 nm) substrates by using magnetron sputtering with the base vacuum better than $8 \times 10^{-9}$ Torr. The Ar pressure during sputtering was 2 mTorr. The dc power for sputtering FGT and Pt was 15 W and the corresponding deposition rates for FGT and Pt were 0.17 Å/s and 0.83 Å/s, respectively. The purity of Fe$_3$GeTe$_2$ target is 99.9% and Pt target is 99.95%. The actual atomic percentages of Fe, Ge, and Te in sputtered FGT films are 42.1%, 24.6%, and 33.3%[34], respectively, in which the Fe concentration is less while Ge is higher than their concentrations in FGT targets. The deposited multilayers were patterned into a Hall bar structure with the length of 50 μm - 200 μm and width of 2 μm - 10 μm by using standard photolithography and ion milling processes.

*Structure characterization*: XRD spectra and X-ray reflectivity (XRR) were recorded on a Bruker D8 X-ray diffractometer. The cross-section and element distribution of FGT/Pt samples were analyzed by using a HRTEM (FEI Titan Themis 200) equipped with an energy-dispersive x-ray spectroscopy (EDS). The TEM samples were prepared by using a focused ion beam system (FIB, FEI-Helios 450S).

*Electrical and magnetic measurements*: All electrical measurements including the angle dependence were performed in a physical property measurement system (PPMS, Quantum Design) in the temperature range of 2 K - 300 K. The magnetization loops were also measured by the PPMS equipped with a vibrating sample magnetometer (VSM).

## Supporting Information

Supporting Information is available free of charge at .

Estimated temperature dependence of spin diffusion length; roughness estimated thorough XRR measurement; comparison of Hall results between Pt/FGT and FGT/Pt; First-principles calculation of $T_c$ in crystalline Pt/FGT bilayers.

**Acknowledgments**

This work is supported by the National Key R&D Program of China (Grants No. 2019YFB2005800, No. 2022YFA1405100, and No. 2018YFA0701500), the National Natural Science Foundation of China (Grant No. 61974160, 61821091, and 61888102), and the Strategic Priority Research Program of the Chinese Academy of Sciences (Grant No. XDB44000000).

Supporting Information

# Proximity-induced interfacial room-temperature ferromagnetism in semiconducting $Fe_3GeTe_2$

Qianwen Zhao[1,2], Yingmei Zhu[3,4], Hanying Zhang[1,2], Baiqing Jiang[1,2], Yuan Wang[1,2], Tunan Xie[1,2], Kaihua Lou[1,2], ChaoChao Xia[1,5], Hongxin Yang[3,4*], and C. Bi[1,2,5*]

[1]Laboratory of Microelectronic Devices & Integrated Technology, Institute of Microelectronics, Chinese Academy of Sciences, Beijing 100029, China

[2]University of Chinese Academy of Sciences, Beijing 100049, China

[3]National Laboratory of Solid State Microstructures, School of Physics, Collaborative Innovation Center of Advanced Microstructures, Nanjing University, Nanjing 210093, China

[4]Ningbo Institute of Materials Technology and Engineering, Chinese Academy of Sciences, Ningbo 315201, China

[5]School of Microelectronics, University of Science and Technology of China, Hefei 230026, China

*hongxin.yang@nju.edu.cn
*clab@ime.ac.cn

**Table of contents**
1. **Temperature dependence of estimated $\lambda_{Pt}$**
2. **Roughness estimated through X-ray reflectivity**
3. **Comparison of Hall results between FGT/Pt and Pt/FGT**
4. **First-principles calculation of $T_c$ in crystalline Pt/FGT bilayers**

# 1. Temperature dependence of estimated $\lambda_{Pt}$

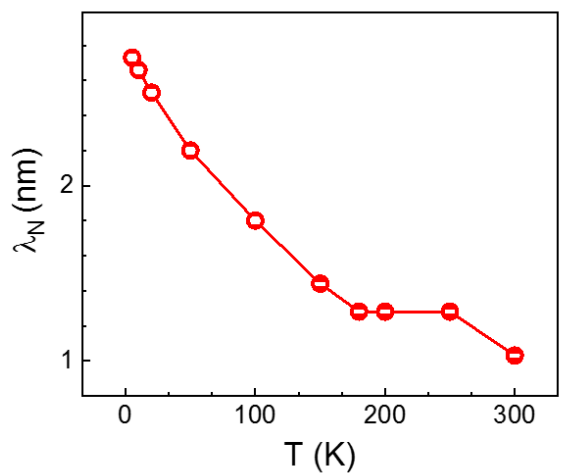

Figure S1. The calculated $\lambda_{Pt}$ as a function of temperature with $\theta_{SH} = 0.097$.

# 2. Roughness estimated through X-ray reflectivity

The roughnesses of Sub/FGT (10)/Pt (3) and Sub/Pt (3)/FGT (10) were also characterized through XRR measurement. Figure S2 shows the reflected X-ray intensity as a function of 2θ, where the roughness-related decay rates are almost the same for both samples. According to these results, the estimated roughnesses for each layer are: FGT 0.29 nm, Pt 0.36 nm in Sub/FGT (10)/Pt (3); FGT 0.44 nm, Pt 0.54 nm in Sub/ Pt (3)/FGT (10).

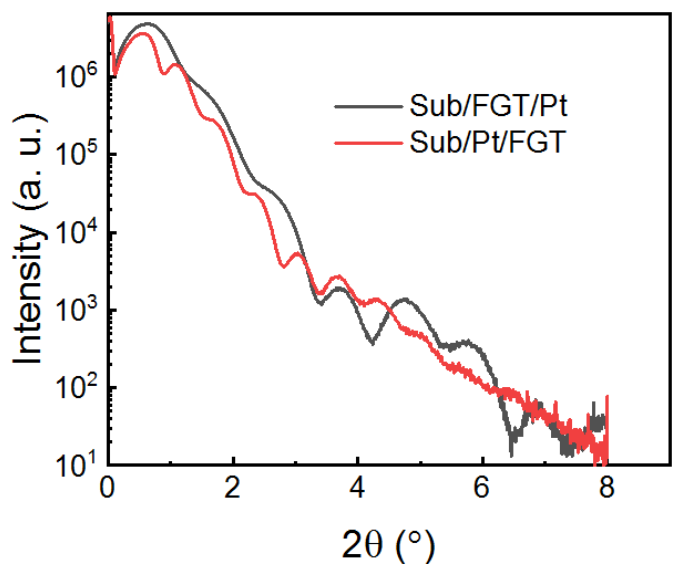

Figure S2. XRR results of Sub/FGT (10)/Pt (3) and Sub/Pt (3)/FGT (10) samples.

# 3. Comparison of Hall results between FGT/Pt and Pt/FGT

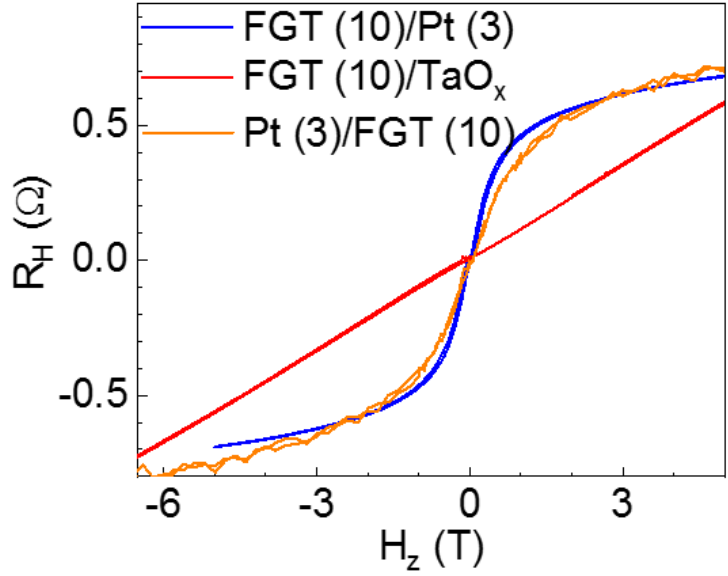

Figure S3. Comparison of Hall measurement results between Sub/FGT (10)/Pt (3) and Sub/Pt (3)/FGT (10).

## 4. First-principles calculation of $T_c$ in crystalline Pt/FGT bilayers

To reveal the possible origin of room-temperature ferromagnetism at the Pt/FGT interfaces, we performed first-principles calculation by using the Vienna ab initio simulation package (VASP) with the generalized gradient approximation (GGA) and projector augmented wave (PAW) potentials[1-4]. The cutoff energy of 400 eV and the K-point mesh of $15 \times 15 \times 1$ were used. We considered a crystalline FGT monolayer, as shown in Figure S4a. $J_z$, $J_{top}$, and $J_{bot}$ represent the exchange coupling between two $Fe^{3+}$, $Fe^{2+}$ and top $Fe^{3+}$, and $Fe^{2+}$ and bottom $Fe^{3+}$, respectively. The top $Fe^{3+}$ will be influenced by adsorbed Pt atoms for calculating the interfacial magnetism as discussed below. As shown in Figure S4b, we calculated the total energy in four magnetic configurations:

$$E_{FM} = -\frac{1}{2} \times \frac{2}{3} \times 12 \times S_1^2 \times J_z - \frac{1}{2} \times 12 \times S_1 S_2 \times \left[\frac{1}{3} \times 3J_{top} + \frac{1}{3} \times 3J_{bot} + \frac{1}{3} \times 3(J_{top} + J_{bot})\right] + E_{other};$$

$$E_{AFM1} = -\frac{1}{2} \times \frac{2}{3} \times 12 \times S_1^2 \times J_z - \frac{1}{2} \times 12 \times S_1 S_2 \times \left[\frac{1}{3} \times J_{top} + \frac{1}{3} \times J_{bot} + \frac{1}{3} \times (J_{top} + J_{bot})\right] + E_{other};$$
(S1)

$$E_{AFM2} = -\frac{1}{2} \times \frac{2}{3} \times 12 \times S_1^2 \times (-J_z) - \frac{1}{2} \times 12 \times S_1 S_2 \times \left[\frac{1}{3} \times 3J_{top} - \frac{1}{3} \times 3J_{bot} + \frac{1}{3} \times 3(J_{top} - J_{bot})\right] + E_{other};$$
(S2)

$$E_{AFM3} = -\frac{1}{2} \times \frac{2}{3} \times 12 \times S_1^2 \times (-J_z) - \frac{1}{2} \times 12 \times S_1 S_2 \times \left[-\frac{1}{3} \times 3J_{top} + \frac{1}{3} \times 3J_{bot} + \frac{1}{3} \times 3(-J_{top} - J_{bot})\right] + E_{other},$$
(S4)
24

where $S_1$ and $S_2$ are the spin operators on site $Fe^{3+}$ and $Fe^{2+}$, respectively.

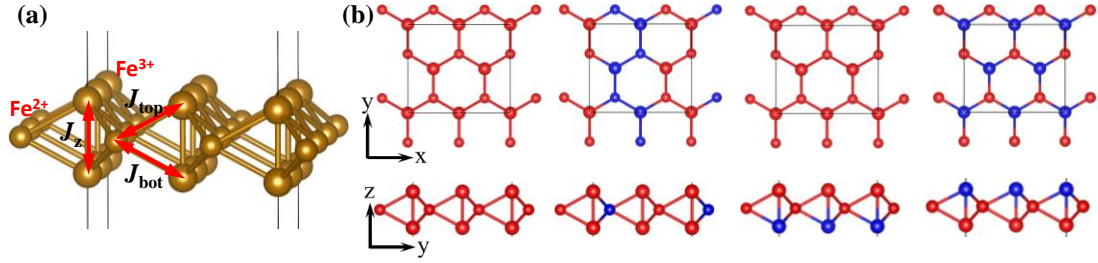

Figure S4. a) Perspective view of Fe atoms of crystallize FGT monolayer. Larger and smaller size balls represent the $Fe^{3+}$ and $Fe^{2+}$, respectively. b) Schematics of magnetic configurations, including one ferromagnetic (FM) and three antiferromagnetic (AFM1, AFM2, and AFM3) configurations.

To calculate the interfacial magnetism induced by adjacent Pt layers, we assume that Pt atoms are adsorbed on the top of either $Fe^{3+}$ or Ge sites, as schematically shown in Figure S5a. Table S1 shows the calculated distance ($d$) between Fe atoms and exchange coupling after all structures are fully relaxed and achieve the most stable states. For the isolated FGT monolayer, $d_{top} = d_{bot}$ and $J_{top} = J_{bot}$ are expected, while with adsorbed Pt atoms, both $d_{top}$ and $d_{bot}$ change and $J_{top}$ increases about two times. In particular, $J_{top}$ is strongest for the Pt atoms adsorbed on the top of Ge. Figure S5b shows the calculated magnetization as a function of temperature, in which the $T_c$ of FGT increases from 82 K to 90 K and 106 K when Pt atoms are adsorbed on the top of $Fe^{3+}$ and Ge sites, respectively. Although these results do not directly show that $T_c$ of FGT monolayer cannot be increased to room-temperature by Pt adsorption, it is demonstrated that Pt atoms indeed increase exchange coupling between two Fe ions and lead to the increase of $T_c$. For the sputtered FGT with similar short-range atomic orders as crystallize FGT, the enhancement of exchange coupling and thus $T_c$ at the FGT/Pt interfaces are also expected.

Table S1. Calculated distance between Fe atoms and exchange coupling for FGT monolayer with/without Pt adsorption.

|  | $d_{top}$ (Å) | $d_{bot}$ (Å) | $d_z$ (Å) | $J_{top}$ (meV) | $J_{bot}$ (meV) | $J_z$ (meV) | $K$ (meV/atom) |
|---|---|---|---|---|---|---|---|
| FGT | 2.62 | 2.62 | 2.49 | 7.96 | 7.96 | 27.04 | 0.83 |
| FGT/Pt (Fe-top) | 2.56 | 2.65 | 2.41 | 12.04 | 6.91 | 12.69 | 0.60 |
| FGT/Pt (Ge-top) | 2.58 | 2.63 | 2.45 | 15.18 | 7.45 | 18.11 | 0.47 |

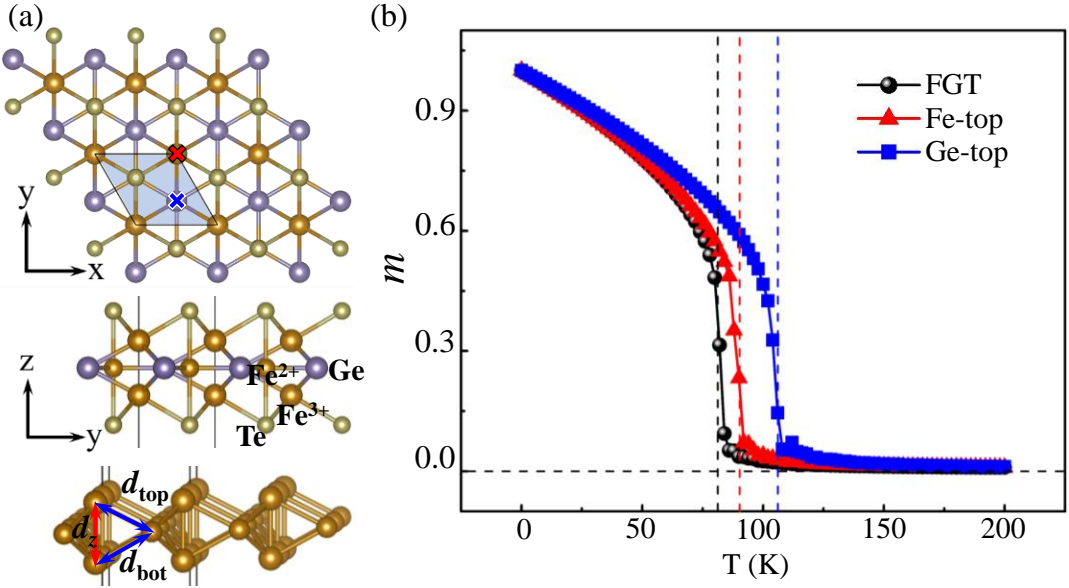

Figure S5. a) Top and side views of FGT monolayer. Red and blue symbol in top parallel of (a) represent the adsorption site of Pt atom. b) The calculated $T_c$ of FGT, FGT/Pt (Fe-top), and FGT/Pt (Ge-top site) is 82 K, 90 K, and 106 K, respectively.